\documentclass[pre,amsmath,superscriptaddress,draft,nofootinbib]{revtex4}

\usepackage{graphicx}
\usepackage{bm}

\begin{document}

\title{Reply to comments by D.H.E.\ Gross}

\author{Christopher Jarzynski}

\affiliation{Theoretical Division, T-13, MS B213, Los Alamos National Laboratory,
Los Alamos, New Mexico 87545 \\
{\tt chrisj@lanl.gov}
}


\maketitle

In a recent posting to these archives~\cite{gross},
Gross asserts that the nonequilibrium work theorem,
\begin{equation}
\label{eq:nwt}
\Bigl\langle \exp(-W/T)\Bigr\rangle = \exp(-\Delta F/T),
\end{equation}
is incorrect. 
He discusses an apparent counterexample, suggesting that it
points to flaws in the derivation of Eq.\ref{eq:nwt}.
He then argues that these flaws originate
in the misuse of Clausius's definition of entropy.
My aim here is to provide a brief response to Gross's analysis.

In Eq.\ref{eq:nwt}, $W$ refers to the work performed on a system
driven away from equilibrium
by the variation of an externally controlled work parameter;
the angular brackets denote an average over infinitely many realizations
of this process;
$T$ is the temperature at which the system is initially prepared
($k_B=1$);
and $\Delta F$ is the free energy difference between two equilibrium
states, corresponding to the initial and final values of the work
parameter.
For a sampling of derivations of Eq.\ref{eq:nwt} for 
classical systems, see 
Refs.~\cite{CJ97a,Crooks98,HummerSzabo01,Evans03,CJ04,ParkSchulten04};
for a broader discussion of theoretical and experimental aspects
of fluctuation theorems, see Ref.~\cite{physToday}.

The counterexample mentioned in Ref.~\cite{gross}, and attributed
to Sung~\cite{sung}, involves the sudden expansion of an ideal gas.
Imagine a thermally isolated, closed container divided by a perfectly
thin but impenetrable partition into
two equal compartments, and imagine that one compartment
initially contains an ideal gas of $N$ particles in equilibrium at temperature $T$,
while the other compartment is empty.
At time $t=0$, the partition is removed, and 
the gas subsequently expands to fill the entire container.
The free energy difference for this process is
\begin{equation}
\label{eq:dF}
\Delta F = -N T \ln 2.
\end{equation}
Since no work is performed in removing the partition ($W=0$),
we seemingly have a contradiction with Eq.\ref{eq:nwt}.

As pointed out by Crooks~\cite{GECthesis},
the discrepancy arises here because the gas is not initially described
by the canonical distribution corresponding to the initial Hamiltonian.
We can write this Hamiltonian as follows:
\begin{equation}
H_0(\Gamma) = \sum_{n=1}^N \Bigl[\frac{p_n^2}{2m} + U({\bf r}_n)\Bigr]
\qquad,\qquad
\Gamma = ({\bf r}_1,{\bf p}_1;\cdots {\bf r}_N,{\bf p}_N)
\end{equation}
(notation hopefully obvious).
The single-particle potential energy function $U({\bf r})$ is zero
within either compartment, and infinite outside the container.
In the corresponding canonical distribution $p\propto\exp(-H_0/T)$,
most microstates $\Gamma$ are characterized by approximately equal numbers of
particles in the two compartments,
rather than by a state in which all $N$ particles reside in one compartment.
Since the derivations of Eq.\ref{eq:nwt} given in
Refs.~\cite{CJ97a,Crooks98,HummerSzabo01,Evans03,CJ04,ParkSchulten04}
all rely strongly on the assumption that
the initial phase space conditions of the system are sampled canonically
with respect to the initial Hamiltonian, and
since the example described in the previous paragraph violates
this assumption, we cannot expect Eq.\ref{eq:nwt}
to be valid in this situation.
(If we do sample our initial conditions from a canonical ensemble --
so that particles are distributed with equal probability in both
compartments -- then $\Delta F=0$, and Eq.\ref{eq:nwt} is satisfied~\cite{GECthesis}.)

This resolution might seem unsatisfying --
what good is Eq.\ref{eq:nwt} if it doesn't apply to such
a classic example of a thermodynamic process? --
but in fact it underscores the 
important point that a canonical sampling of initial conditions is a necessary
condition for the validity of Eq.\ref{eq:nwt}.

The removal of a partition is not the only way to achieve the expansion of a gas.
An alternative approach involves the familiar piston-and-gas construction,
in which the container holding the gas is closed off at one
end by a piston whose position is controlled externally.
Imagine that the gas is first prepared in thermal equilibrium, with
the piston fixed at some location $A$;
then the piston is withdrawn -- at a speed $u$ much greater than
the thermal speed $v_{\rm th}=\sqrt{3T/m}$ of the gas particles -- to a new location $B$,
doubling the volume of the container.
The piston is then held fixed as the gas expands into the newly available volume.
As before, $\Delta F = -N T \ln 2$.
Moreover, during a {\it typical} realization of this process,
not a single gas particle strikes the moving piston (since $u\gg v_{\rm th}$),
therefore $W=0$.
Once again there is an apparent contradiction with Eq.\ref{eq:nwt}.
However, as Lua and Grosberg~\cite{lg} have shown with explicit calculations,
the average of $\exp(-W/T)$ in this case is dominated by {\it atypical}
realizations for which one or more particles begin with
speeds sufficiently large that they do in fact collide with the rapidly moving
piston.
The likelihood of such realizations decreases severely with piston speed,
but this is compensated by a large energy exchange
between piston and particle (large value of $-W$),
and Eq.\ref{eq:nwt} remains valid for any finite piston speed $u$;
see Ref.~\cite{lg} for details.
The larger the value of $u$, the more the average of
$\exp(-W/T)$ is dominated by extremely unlikely realizations,
in which at least one particle's velocity is drawn
from deep within the tail of the Maxwellian distribution.

In Ref.~\cite{sung}, Sung also considers the piston-and-gas example
(in one spatial dimension),
and analyzes the situation in which the initial particle speeds
are sampled from a {\it truncated} Maxwellian distribution:
no particle is allowed to have an initial speed greater than $\xi$,
a positive but otherwise arbitrary cutoff value.
Sung obtains an explicit expression for $\langle\exp(-W/T)\rangle$
(Eq.15), which 
converges to $\exp(-\Delta F/T)$ in the limit $\xi\rightarrow\infty$~\cite{sung}.
In other words the validity of Eq.\ref{eq:nwt} is recovered when the initial
velocities are sampled from a {\it full} (rather than truncated) Maxwellian distribution.
This example again confirms the nonequilibrium work theorem,
and highlights the importance of the canonical distribution for its validity.

In discussing these examples, I do not mean to dismiss the genuine
subtleties that arise when applying Eq.\ref{eq:nwt} to
situations involving hard walls, impenetrable partitions,
instantaneously withdrawn pistons, and the like~\cite{private}.
Eq.\ref{eq:nwt} involves an implied limit of infinitely many realizations,
and when other infinities are present
(such as the potential energy associated with a perfectly hard wall)
then -- not surprisingly -- the order in which limits are taken can become
an important issue, requiring careful treatment.

A canonical sampling of initial conditions can generally be achieved by
equilibrating the system with an external heat bath.
Subsequently, we can either maintain thermal contact between the system and
the bath as we vary the work parameter,
or else we can remove the bath before varying the work parameter.
Eq.\ref{eq:nwt} remains valid in either case.
In other words,
while we require a heat bath to generate the canonical initial conditions,
its presence is optional during the remainder of the thermodynamic process.
The piston-and-gas example of Refs.~\cite{sung,lg} illustrates the situation
in which the system is thermally isolated during the process (no heat bath).
By contrast, Ref.~\cite{mj} illustrates the nonequilibrium work theorem
for a model system (a dragged harmonic oscillator) that remains coupled to
a heat bath as the work parameter is varied from its initial to its final value.

While a canonical initial distribution of the system's degrees of freedom is
a necessary condition for the validity of Eq.\ref{eq:nwt},
Adib has recently derived an analogous result,
valid for microcanonically sampled initial conditions~\cite{adib}.

Much of the discussion in Ref.~\cite{gross} focuses not on Eq.\ref{eq:nwt} above, 
but rather on a {\it detailed fluctuation theorem},
\begin{equation}
\label{eq:dft}
\frac
{P_+({\bf z}_B,+\Delta S\vert {\bf z}_A)}
{P_-({\bf z}_A^*,-\Delta S\vert {\bf z}_B^*) } =
 \exp (\Delta S) ,
\end{equation}
derived in Ref.~\cite{dft}.
The two results are only tangentially related:
while Eq.\ref{eq:nwt} can be obtained as a consequence of Eq.\ref{eq:dft},
this represents an indirect route of derivation.
None of the derivations of the nonequilibrium work theorem presented in
Refs.~\cite{CJ97a,Crooks98,HummerSzabo01,Evans03,CJ04,ParkSchulten04}
rely on the validity of the detailed fluctuation theorem.

Note that in Ref.~\cite{dft}, the quantity $\Delta S$ is {\it defined} by the equation
$\Delta S = -\Delta Q/T$
(more precisely, by its generalization to the case of many reservoirs,
Eq.(2) of Ref.~\cite{dft}),
and is explicitly interpreted as the change in the
entropy {\it of the thermal environment}.
By contrast, in Ref.~\cite{gross} the notation $\Delta S$ seems to designate
a different quantity, namely the change in the combined entropy of the environment
{\it and the system}.
This difference in definitions might account for the disagreement between
Refs.~\cite{gross} and \cite{dft}.
(Indeed, the quantity denoted by $\Delta S$ in Ref.\cite{dft} seems to be the
same as the quantity denoted by $\Delta S_b$ in Ref.\cite{gross},
keeping in mind that the two papers use opposite sign conventions for $Q$.)
I believe that Eq.\ref{eq:dft} above is correct
for the quantities as they are defined in Ref.~\cite{dft},
and in any case this issue is not of direct relevance to the validity of Eq.\ref{eq:nwt}.

\end{document}